\documentclass[a4paper,11pt]{article}

\usepackage{jheppub}

\usepackage[T1]{fontenc}

\usepackage{wasysym}

\usepackage{comment}
\usepackage{graphicx}
\usepackage{amsmath}
\usepackage{amsfonts}
\usepackage{amssymb}
\usepackage{mathrsfs}
\usepackage{natbib}
\usepackage{xcolor}
\usepackage[colorlinks=true
,urlcolor=blue
,anchorcolor=blue
,citecolor=blue
,filecolor=blue
,linkcolor=blue
,menucolor=blue
,linktocpage=true
,pdfproducer=medialab
,pdfa=true
]{hyperref}
\usepackage{color}
\usepackage{comment}
\usepackage{xcolor}

\usepackage{cleveref}

\newcommand{\bl}{\left}
\newcommand{\br}{\right}

\newcommand{\Fig}[1]{Fig.~\ref{#1}}
\newcommand{\Eq}[1]{Eq.~(\ref{#1})}

\newcommand{\beq}{\begin{equation}}
\newcommand{\eeq}{\end{equation}}
\newcommand{\bea}{\begin{eqnarray}}
\newcommand{\eea}{\end{eqnarray}}
\newcommand{\bef}{\begin{figure}}
\newcommand{\eef}{\end{figure}}

\newcommand{\AU}{\mathrm{A.U.}}

\newcommand{\MJupiter}{m_\text{\small \jupiter}}
\newcommand{\RJupiter}{R_\text{\small \jupiter}}
\newcommand{\MJuno}{m_\text{\tiny Juno}}

\newcommand{\Domega}{\langle \Delta \omega \rangle}

\begin{document}

\title{\Large The Juno Mission as a Probe of Long-Range New Physics}

\author[a]{Praniti Singh,}
\author[a]{Shi Yan,}
\author[a]{Itamar J.~Allali,}
\author[a,b]{JiJi Fan,}
\author[a]{Lingfeng Li}


\affiliation[a]{Department of Physics, Brown University, Providence, RI 02912, USA}
\affiliation[b]{Brown Theoretical Physics Center, Brown University, Providence, RI 02912, USA]}

\emailAdd{praniti\_singh@brown.edu}
\emailAdd{shi\_yan@brown.edu}
\emailAdd{itamar\_allali@brown.edu}
\emailAdd{jiji\_fan@brown.edu}
\emailAdd{lingfeng\_li@brown.edu}


    \begin{abstract}{
   Orbits of celestial objects, especially the geocentric and heliocentric ones, have been well explored to constrain new long-range forces beyond the Standard Model (SM), often referred to as fifth forces. In this paper, for the first time, we apply the motion of a spacecraft around Jupiter to probe fifth forces that don't violate the equivalence principle. The spacecraft is the Juno orbiter, and ten of its early orbits already allow a precise determination of the Jovian gravitational field. We use the shift in the precession angle as a proxy to test non-gravitational interactions between Juno and Jupiter. Requiring that the contribution from the fifth force does not exceed the uncertainty of the precession shift inferred from data, we find that a new parameter space with the mass of the fifth-force mediator around $10^{-14}$ eV is excluded at 95\% C.L.       }
    \end{abstract}
    \maketitle

\section{Introduction}

The gas giant Jupiter, the largest planet in our solar system and fifth one from the Sun, has been observed extensively for much of human history. With a total of eleven space missions sent from Earth to Jupiter so far, Jupiter has been an important topic of study to learn more about planetary science and other branches of astrophysics. The most recent Jupiter mission, the spacecraft Juno \cite{2017SSRv..213....5B}, was launched in 2011 and has since orbited Jupiter more than 60 times, gathering extensive data with which to study the planet. In addition to learning directly the properties of Jupiter, recent work \cite{Li:2022wix,Yan:2023kdg} has shown the unexpected utility of Juno's data for probing physics beyond the standard model of particle physics (BSM).\footnote{See \cite{Leane:2021tjj,French:2022ccb,Blanco:2023qgi,Linden:2024uph,Ansarifard:2024fan,Blanco:2024lqw} for the use and proposed use of other Jupiter data to study new physics.} 

One unexplored avenue to study BSM physics using Juno data is in the understanding of the gravitational interactions between Juno and Jupiter. The precise measurements of Juno's orbital motion, accounting for all known disturbances by nearby objects, has been used to characterize the properties of Jupiter's gravitational field \cite{durante2020jupiter}. Thus, deviations of Juno's orbit from the expected behavior could be a response to additional forces acting between Juno and Jupiter. Therefore, the data of Juno's orbits, and the degree to which they are explained by Jupiter's gravitational influence, can be a powerful probe of an interaction arising from BSM physics, sometimes referred to as a ``fifth force."\footnote{Strictly speaking, the Higgs-mediated force should be counted as the fifth force. Here we take a traditional point of view calling a force mediated by a BSM particle a fifth force.}

\begin{figure}
    \centering
    \includegraphics[width=0.7\linewidth]{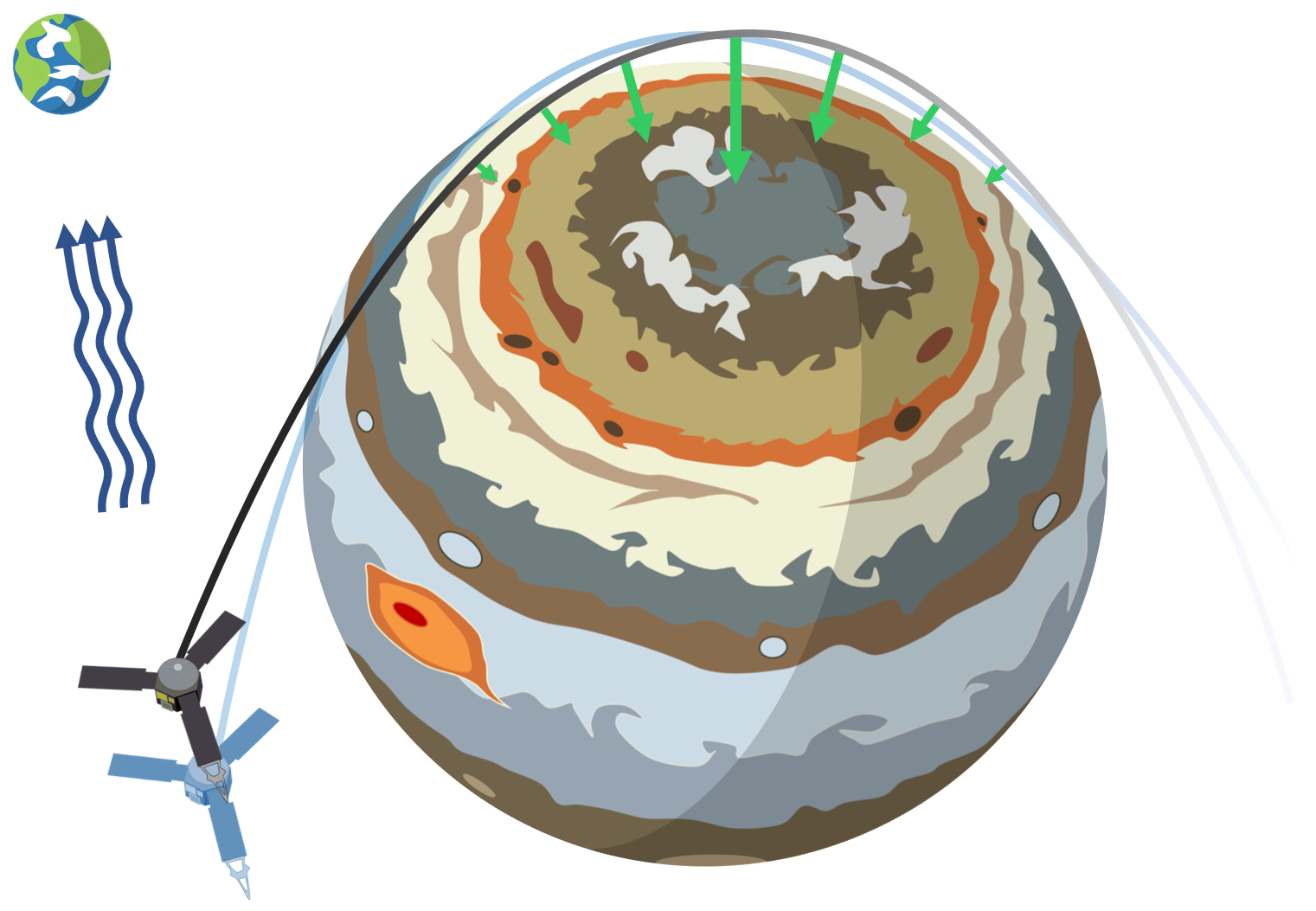}
    \caption{A schematic which highlights the highly eccentric orbit of Juno around Jupiter, as well as its very near approach to Jupiter's surface. Here, the green arrows and their lengths stand for the fifth force that Juno would experience along its trajectory. When the range of the fifth force is small, the fifth force effect is only significant near Juno's closest approach to Jupiter (or perijove) and changes drastically with time. Consequently, Juno's orbit may deviate from our expectations, as shown in the light blue shaded trajectory.}
    \label{fig:schematic}
\end{figure}

Fifth forces can arise from a variety of new physics scenarios. Broadly, we consider at least a new particle, typically a light vector or scalar, which would mediate a Yukawa-type interaction with a finite range related to the mass of the new particle. The range can be quite large for a light mediator, and thus one would expect that orbital dynamics of a celestial body or spacecraft would be sensitive to such an interaction. In \cref{section: fifth}, we provide a precise parameterization for this class of fifth forces. Examples of possible BSM models leading to a long-range Yukawa interaction include gauged $U(1)_{B}$~\cite{Carone_1995,FileviezPerez:2010gw}, $U(1)_{B-L}$~\cite{PhysRevD.20.776,PhysRevLett.44.1316,PhysRevLett.59.393}, and $L_\mu-L_{e}$~\cite{Escudero:2019gzq,doi:10.1142/S0217732391000543,PhysRevD.44.2118}, as well as scalars coupled to baryons~\cite{Blinov:2018vgc,Sibiryakov_2020,Izaguirre_2015,Pospelov_2018}.\footnote{A similar form of interaction also arises in Yukawa gravity theories~\cite{1981Natur.291..636G,Capozziello:2009vr}; the constraints in our work can be applied to these scenarios. }

Geocentric and heliocentric orbital data has been used extensively in the past to apply constraints on the strength and range of fifth forces \cite{Poddar_2021,konopliv2011mars,Williams_2004,Murphy_2013,1985JGR....90.9221S,1987GeoRL..14..730R,Tsai:2021irw,Tsai:2023zza}. In this work, we expand this methodology by utilising, for the first time, the orbit of an object around Jupiter. 
We will make use of the harmonic expansion of Jupiter's gravitational field in \cite{durante2020jupiter}, which describes the deviation of Jupiter's gravitational potential from spherical symmetry and accounts for modeled astrophysical effects in the vicinity of the Juno-Jupiter system.
The uncertainty in these gravitational parameters is used to quantify the uncertainty in the precession of Juno's orbit. Then, to
provide constraints on the strength and range of the fifth force, we require that additional precession induced by the fifth force does not exceed the quantified uncertainty. Juno's orbit is highly eccentric ($1-e \sim 0.01$), and its perijoves or points of closest approach to Jupiter are on the order of $0.05 \,\RJupiter$, where $\RJupiter$ is the radius of Jupiter, as represented in \cref{fig:schematic}. For this reason, Juno's orbits should be sensitive to deviations arising from a fifth force, particularly in a regime of length scales corresponding to Juno's orbit. We will thus show that via this method, the Juno data indeed sets the strongest constraint in this window of parameter space, for a fifth force which does not maximally violate the equivalence principle.

The paper is organized as follows. In \cref{section: fifth}, we provide the phenomenological parameterization of fifth forces that we will study in this work. Then we will compute the predicted shifts in precession angle as a function of fifth-force parameters. In \cref{section: method}, we will describe the method of quantifying the uncertainty in Juno's orbital motion from the gravitational data. In \cref{section: results}, we will show the constraints on the strength of the fifth force as a function of the force range, by requiring that the fifth force causes orbital shifts not greater than the uncertainty from the data. Finally, in \cref{section: conclusions}, we will provide concluding remarks and discuss several future directions.

\section{Fifth force and orbital precession}\label{section: fifth}


As briefly reviewed in the introduction, a fifth force could arise from a BSM scenario with a new light particle, either a scalar or a vector, serving as the force mediator~\cite{Fischbach:1996eq,Fischbach1998}. This interaction is often taken to be of the Yukawa type, associated with a potential energy between two point objects of masses $M$ and $m$ as
\begin{equation}
U(r) \bigg|_{\textrm {point objects}}=-\alpha \frac{G M m}{r} e^{-\frac{r}{\lambda}}~,
 \label{equ:fifth_force_potential}
\end{equation}
where $G$ is the gravitational constant, $r$ is the radial separation between the two masses, $\lambda$ is the fifth-force range, and $\alpha \ll 1$ is a dimensionless parameter characterizing the strength of the fifth force. The mass of the fifth-force mediator, $m_*$, is related to the force range as $m_* \equiv \hbar c / \lambda$. $\alpha$ is determined by the parameters in the underlying UV theory. For instance, if the two masses $M$ and $m$ carry net charges $Q$ and $q$ under the fifth force respectively,  
\begin{equation}
\alpha= \pm \frac{g^2}{4 \pi G} \frac{Q q}{M m}~,
\end{equation}
where $g$ is the coupling constant of the fifth force. In the equation above, $+$ corresponds to an attractive force and $-$ corresponds to a repulsive one.

The potential energy in \cref{equ:fifth_force_potential} leads to further deviation of an object's orbit around a massive celestial body beyond that due to Newtonian perturbations and corrections from general relativity (GR). The deviation can be quantified using orbital parameters or orbital elements as given in \cref{appendix:orbital_elements}. In this work, we study the system of the Juno spacecraft~\cite{durante2020jupiter} orbiting around Jupiter, focusing on the deviation of the argument of periapsis $\omega$, one of the orbital elements listed in \cref{appendix:orbital_elements}. This quantity is understood as the precession of Juno's closest approach to Jupiter about the axis normal to Juno's orbit; we will refer to $\omega$ most often as the ``precession angle" in this work. In addition to the fifth force, GR effects and the non-spherical shape of Jupiter also contribute to deviations in the orbital precession of Juno. These non-BSM contributions will be explained in detail in the next \cref{section: method}.

 The potential energy in \cref{equ:fifth_force_potential} assumes point-like objects. In practice, since Jupiter has a non-trivial density profile, the potential energy between Jupiter and Juno will be corrected as
\begin{equation}
U(r)= -\int_{\textrm{V}} \left(\alpha \frac{G \MJuno \rho(r')}{|\Vec{r}-\Vec{r'}|} \times e^{-\frac{|\Vec{r}-\Vec{r'}|}{\lambda}}\right) d^3 \Vec{r'}
 ~,
 \label{equ:fifth_force_2layer}
\end{equation}
where $\rho(r)$ denotes the mass density of Jupiter\footnote{Here we assume that the charge density of Jupiter is proportional to the mass density up to an overall factor, which is captured in $\alpha$. This applies to some models such as $U(1)_B$, but for other models, e.g., $U(1)_{B-L}$, the computation needs to be modified correspondingly.  } and the integration is performed over the volume of Jupiter. In this work, we use the 2-layer density profile of Jupiter $\rho(r)$ given in Ref.~\cite{Militzer_2024},
while Juno is still considered a point object with a mass $\MJuno$.

In order to derive the non-periodic drift of $\omega$, we use the orbital perturbation theory~\cite{milani1987non} and treat the fifth force as a radial perturbation at leading order. The rate of orbital precession due to the fifth force is given by the following Gauss planetary equation:
\begin{equation}
    \dot{\omega} = -\sqrt{\frac{a (1 - e^2)}{e^2 \mu}} \frac{1}{\MJuno} \frac{\partial U}{\partial r} \cos f,
    \label{equ:dot_omega}
\end{equation}
where $\mu\equiv G\MJupiter$, with $\MJupiter$ the mass of Jupiter, $a$ is the semi-major axis, $e$ is the eccentricity, $f$ is the phase angle with respect to the perijove (also called true anomaly) of the Newtonian orbit, and $U$ is the corrected potential energy in \cref{equ:fifth_force_2layer}. The corresponding averaged precession deviation per orbit is given by:
\begin{equation}
    \Domega = \frac{1}{2\pi} \int_0^{2 \pi} \frac{d \omega}{d f} \; d f = \frac{1}{2\pi} \int_0^{2 \pi} \frac{\dot{\omega}}{h / r^2} \; d f~, \label{eq:deltaomega_fifth}
\end{equation}
where $h$ is the angular momentum per unit mass and $r(f)=a(1-e^2)/(1+e\cos f)$ is Juno's distance from Jupiter's center.

\begin{figure}
    \centering
    \includegraphics[width=0.48\textwidth]{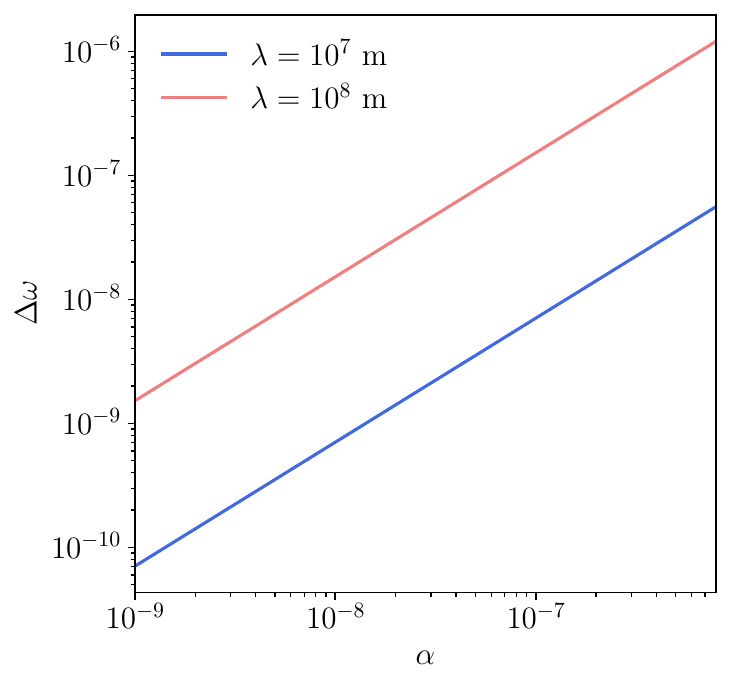}
    \includegraphics[width=0.48\textwidth]{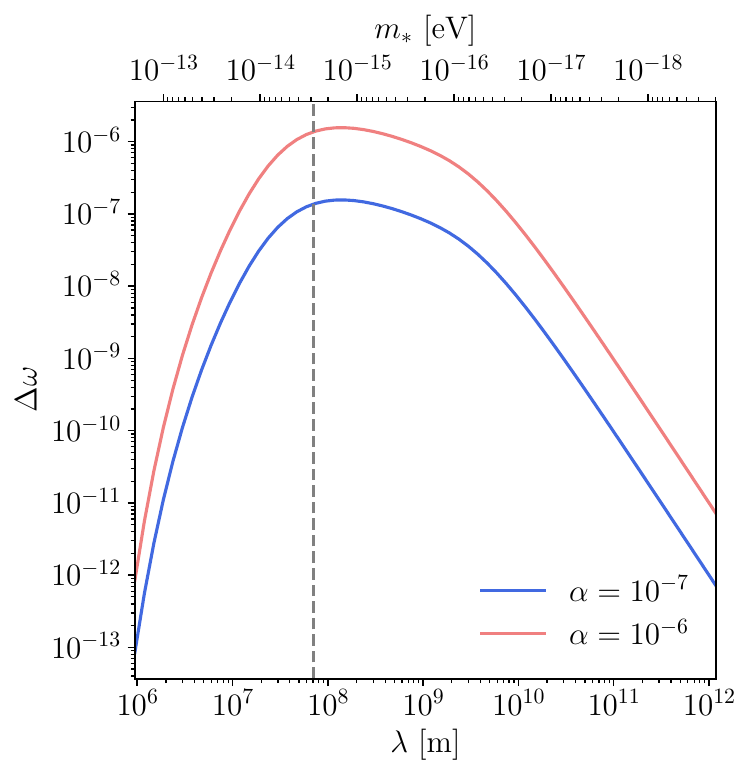}
    \caption{Left: $\Delta \omega$ as a function of $\alpha$ when $\lambda$ is fixed at $10^{-1}~\AU$ and $10^{-2}~\AU$ Right: $\Delta \omega$ as a function of $\lambda$ when $\alpha$ is fixed to $10^{-7}$ and $10^{-6}$; the vertical grey dashed line denotes $\lambda = \RJupiter$.}
    \label{fig:scan_demo}
\end{figure}

We plot $\Domega$ as a function of $\alpha$ for a given $\lambda$ in the left panel of~\cref{fig:scan_demo};
one can see that $\Domega$ increases with $\alpha$ linearly, as the strength of the fifth force increases. In the right panel of~\cref{fig:scan_demo}, we show $\Domega$ as a function of $\lambda$ for a given $\alpha$. In this case, $\Domega$ first increases and then decreases, with a peak at $\lambda \approx 2.0 \times 10^{-4}~\AU \approx 0.4~\RJupiter$, where $\AU\sim 1.5\times 10^{11}$~m stands for astronomical units and  $\RJupiter \approx 7.1 \times 10^4$~km is the Jovian radius.
The minimum value of $\Domega$ is achieved at both the low and high mass ends, which aligns with our expectations. At short ranges $(\lambda \ll0.4~\RJupiter)$, the fifth force is only effective at distances shorter than the orbital size, and thus has little effect on Juno. On the other hand, when $\lambda$ increases above 0.4 $\RJupiter$, the fifth force affects motion at scales above Juno's orbit and the induced precession becomes negligible. 
Both limits result in negligible shifts in the precession angle.


\section{Juno's precession from data}\label{section: method}

Jupiter's gravitational field has been studied using Juno's orbits in \cite{2017GeoRL..44.4694F,2018AGUFM.P23A..04P,durante2020jupiter}, reported as measurements of the mass and harmonic expansion coefficients (characterizing Jupiter's deviation from being a perfect sphere). In this section, we will discuss how to estimate the uncertainty of the drift in Juno's precession angle from the derived gravity field. 

The effective gravitational potential consisting of Newtonian gravity and the leading relativistic correction can be expanded as~\cite{capderou2005satellite} 
\begin{equation}
    V (r, z) = \frac{\mu}{r} \bl[ 1 + \frac{\mu a (1 - e^2)}{c^2 r^2} - \sum^\infty_{l=2} \bl( \frac{\RJupiter}{r} \br)^l J_l P_l\left( \frac{z}{r} \right) \br]~,
    \label{equ:reduced_potential}
\end{equation}
where $c$ is the speed of light, $P_l$'s are the Legendre polynomials, $J_l$'s are the corresponding Legendre coefficients, and $z$ is Juno's distance to the Jovian equatorial plane and is taken to be positive when Juno is located north of the plane. There are three terms in the brackets. The first term is the leading Newtonian potential. The second term gives the leading correction to the effective potential from GR. The third term with the $J_l$'s gives the contribution of non-spherical expansion of the Jovian gravity field. Since Jupiter rotates fast, its oblate shape leads to a significant $J_2\sim 0.015$. The other $J_l$ coefficients are much smaller.

The fifth force we study is spherically symmetric at leading order.\footnote{For simplicity, we neglect higher-order non-spherical corrections arising from the shape of Jupiter, which will lead to at most $\mathcal{O}(10^{-2})$ corrections to the constraints derived.} In addition, the observation of Juno lasts much longer than the rotational periods of Jupiter and its major moons. 
Therefore we assume that the non-spherical correction of $V$ is axially symmetric. In other words, further terms in \cref{equ:reduced_potential} with explicit $x,y$ or equivalently longitude dependence are dropped.\footnote{In the literature, the last term in \cref{equ:reduced_potential} is often referred to as ``zonal harmonics", while the other terms are called ``sectorial" and ``tesseral" harmonics according to their latitude dependence.}

 To see how the effective gravitational potential affects the precession angle, we set a coordinate system with three orthogonal coordinates: $\hat{e}_z$ being the unit vector pointing from the south pole to the north pole of Jupiter, $\hat{e}_N$ the unit normal vector of Juno's orbital plane, and $\hat{e}_R$ the unit vector pointing from the Jovian center to Juno. The gravitational force that Jupiter exerts on Juno is then given by
\bea
    \pmb{F} &=& \MJuno\left( - \frac{\partial V}{\partial r} \hat{e}_R + \frac{\partial V}{\partial z} \hat{e}_z \right)\\
    &=& \MJuno\left\{ \bigg[\sin i \sin (\omega + f) \frac{\partial V}{\partial z} - \frac{\partial V}{\partial r}\bigg] \hat{e}_R + \sin i \cos (\omega + f) \frac{\partial V}{\partial z} \hat{e}_T + \cos i \frac{\partial V}{\partial z} \hat{e}_N \right\}~,\nonumber
    \label{equ:force}
\eea
where $i$ is the inclination angle of Juno's orbit. Then, the rate of change of the precession angle due to gravity is~\cite{milani1987non}
\begin{equation}
    \dot{\omega}_g = \sqrt{\frac{a (1 - e^2)}{\MJuno^2 e^2 \mu}} \left\{ - \cos f F_R + \frac{(2 + \cos f) \sin f}{1  + \cos f} F_T - \frac{e \sin (\omega + f)}{\sin i (1 + \cos f)} F_N\right\}~,
    \label{equ:changing_rate_omega}
\end{equation}
where $F_R$, $F_T$, and $F_N$ are the components of the gravitational force given in \Eq{equ:force}, in the $\hat{e}_R$, $\hat{e}_T$ and $\hat{e}_N$ directions, respectively. To distinguish from the shift in precession angle due to the fifth force in \cref{equ:dot_omega}, we denote the gravity-induced rate of change as $\dot{\omega}_g$. 
Integrating over one period, the effect from the leading
spherically-symmetric Newtonian term will be zero. Only the GR term and zonal harmonics contribute to the average value of the shift in $\omega_g$, which is defined as
\begin{equation}
    \bl< \Delta \omega_g \br> = \frac{1}{2 \pi} \int_0^{2 \pi} \frac{\dot{\omega}_g}{h/r^2} \; d f~.
    \label{eq:omega_analytic}
\end{equation}

\begin{figure}
    \centering
    \includegraphics[width=0.54\textwidth]{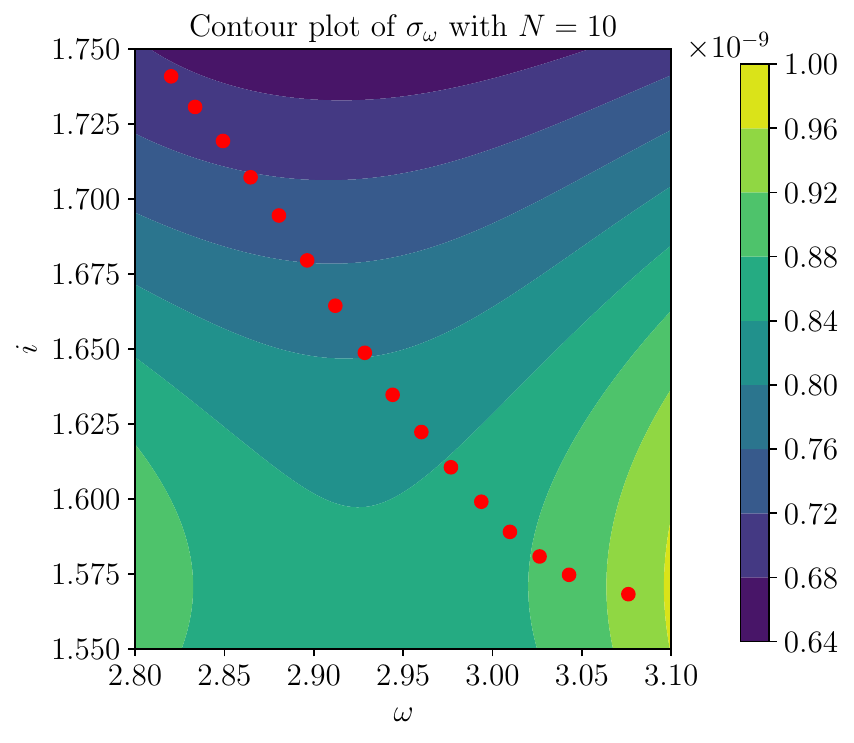}
    \includegraphics[width=0.44\textwidth]{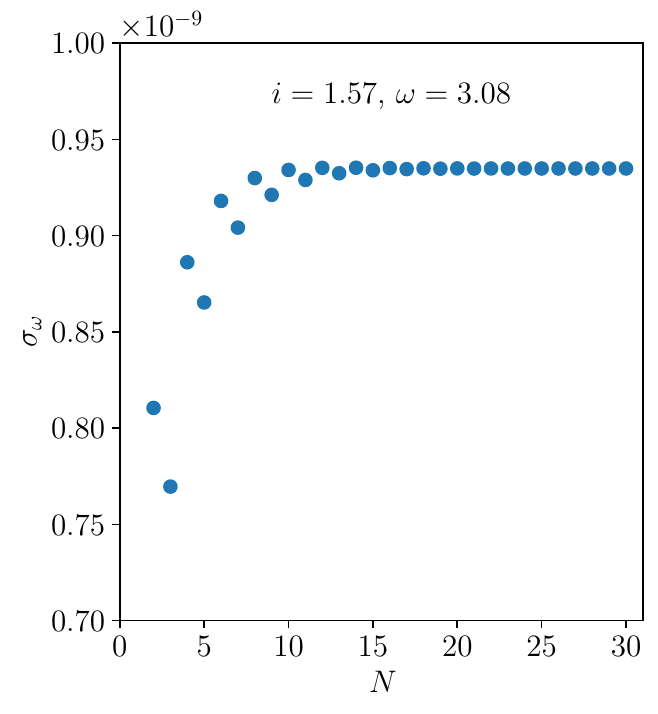}
    \caption{Left: Contour plot of the uncertainty in the measured precession of Juno's orbits, $\sigma_ \omega$, with the highest order of $J_l$ computed $N = 10$ in the $(\omega, i)$ plane. The red dots represent $(\omega, i)$ of the first and third to seventeenth Juno orbits, from right to left. Right: the uncertainty in the measured precession of Juno's orbits, $\sigma_\omega$, plotted against the highest order of $J_l$ computed, $N$, for $i = 1.57$ and $\omega = 3.08$.  }
    \label{fig:PJs_distribution}
\end{figure}

The uncertainty of $\Delta\omega_g$ per perijove is not directly obtainable from the literature. Instead we infer it from the uncertainties of the zonal harmonic coefficients and other parameters. More specifically, the variance (or equivalently the uncertainty of $\Delta\omega_g$) from the Juno data takes the form
\begin{equation}
    \sigma^2_\omega = \frac{\partial \bl< \Delta \omega_g \br>}{\partial \pmb{J}} C_J \left( \frac{\partial \bl< \Delta \omega_g \br>}{\partial \pmb{J}} \right)^T~,
    \label{eq:sigmaomega}
\end{equation}
where $\pmb{J} = [\mu, J_2, J_3, \cdots, J_N]$ is an $N$-dimensional vector of the gravitational field coefficients, $C_J$ is an $N \times N$ covariance matrix provided by \cite{durante2020jupiter} and $\frac{\partial \bl< \Delta \omega_g \br>}{\partial \pmb{J}}$ could be obtained by combining \crefrange{equ:reduced_potential}{eq:omega_analytic}.
We derive each $\frac{\partial \bl< \Delta \omega_g \br>}{\partial \pmb{J}}$ analytically and present these expressions in \cref{app:eqs}.

The orbital data of Juno provided in Ref.~\cite{connerney2017junodata} shows that $i$ and $\omega$ of each orbit are different. The left panel of \Cref{fig:PJs_distribution} is a contour plot of $\sigma_ \omega$ computed up to $N = 10$ in the $(\omega, i)$ plane. The red dots represent $(\omega, i)$ of Juno's first and third through seventeenth
orbits (data from the second orbit was lost) with the rightmost point for the first orbit and the leftmost one for the seventeenth. Note that we base our analysis on only the first seventeen orbits since the analysis of Jupiter's gravitational field has so far been conducted for this subset of the data \cite{durante2020jupiter}. The Juno orbits' $\omega$ spans from 2.8 to 3.1 while $i$ spans from 1.55 to 1.75. The corresponding $\sigma_\omega$ ranges from from $7.1 \times 10^{-10}$ to $9.0 \times 10 ^{-10}$. To set a conservative constraint on the fifth force, we choose $i = 1.57$ and $\omega = 3.08$ of the first orbit, which gives the largest $\sigma_\omega$. 

We then vary $N$ to see how it affects the estimated uncertainty of the precession angle. In the right panel of \cref{fig:PJs_distribution}, we fix $i = 1.57$, $\omega = 3.08$ and plot $\sigma_\omega$ versus $N$. $\sigma_ \omega$ increases a bit from $N = 2$ to $\sim 10$ and then plateaus at $0.9 \times 10^{-9}$. Even though higher $J_l$'s have uncertainties comparable to the lower $J_l$'s, the asymptotic value of $\sigma_\omega$ indicates that the covariance of these quantities is important, while treating them independently would overestimate the error. Furthermore, this is consistent with our approach that the measurements of the $J_l$'s largely stem from measurements of Juno's orbital motion, in this case the change in $\omega$. Thus to be most conservative, we choose $N = 10$ as it gives the largest $\sigma_\omega$, which is $ \approx 9.0 \times 10^{-10}$. In summary, we compute $\sigma_\omega$ in \cref{eq:sigmaomega}, choosing $i = 1.57$, $\omega = 3.08$, and computing up to $J_{10}$.


\section{Results}\label{section: results}

\begin{figure}
    \centering
    \includegraphics[width=\textwidth]{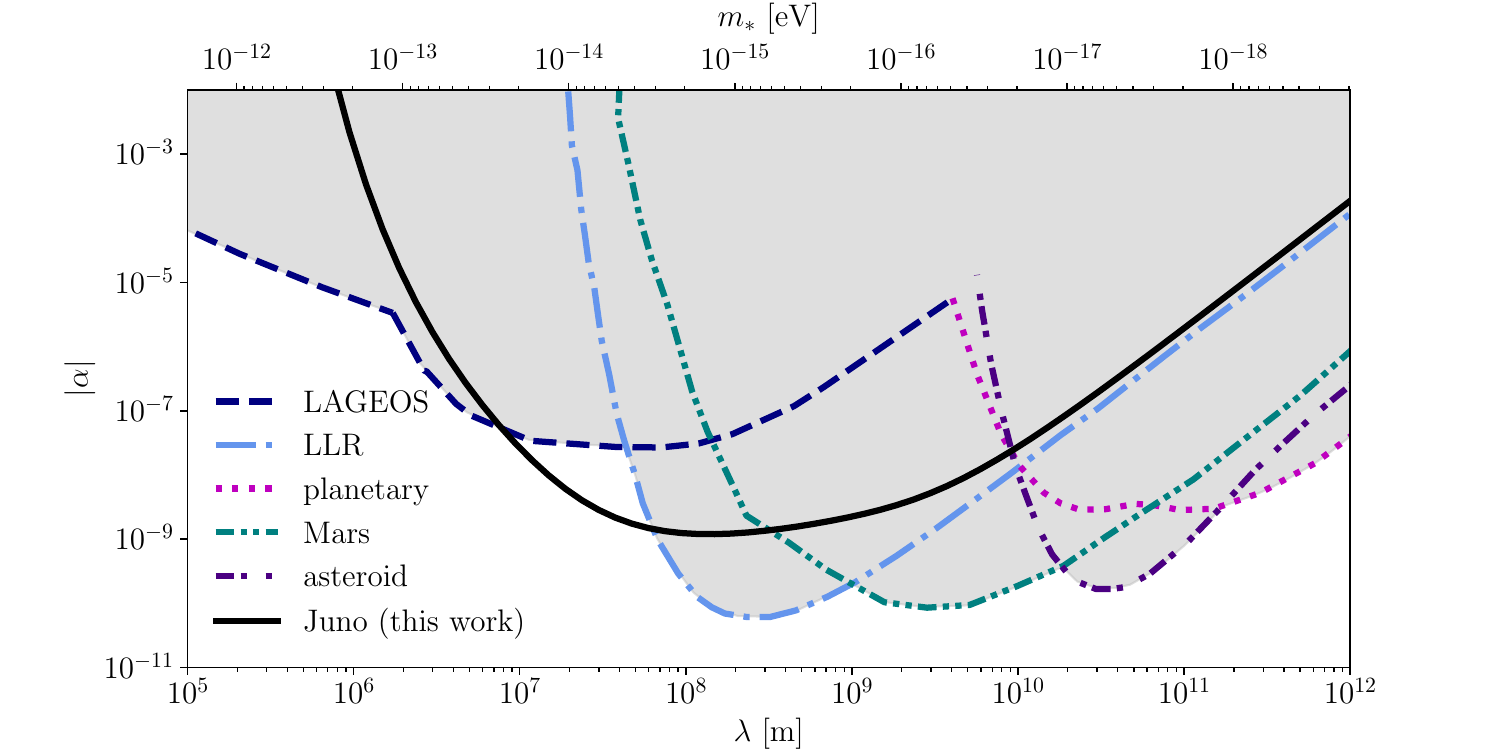}
    \caption{Constraints on the fifth-force parameters $\lambda$ and $\alpha$, where the region above the curves is ruled out (regions constrained previous to this work are shaded in grey). The deep blue dashed line is the constraint using LAGEOS~\cite{1985JGR....90.9221S,1987GeoRL..14..730R}; the light blue dash-dotted line is the constraint from LLR~\cite{Adelberger:2003zx}; the green dense dash-dotted line is from the Mars tracking data analysis~\cite{konopliv2011mars}; the pink dotted line is from the planetary limit~\cite{Poddar_2021}; the purple loose dash-dotted line is from the asteroid limit~\cite{Tsai:2023zza}, and the black solid line is our result using the Juno orbital data.
    }
\label{fig:results}
\end{figure}

To set the constraint at 95\% confidence level (C.L.), we require $ |\Domega|$ in \cref{eq:deltaomega_fifth} to be  $\leq 2 \sigma_\omega$ computed by \cref{eq:sigmaomega}. \Cref{fig:results} shows the corresponding constraint on $\alpha$ characterizing the fifth-force strength as a function of the fifth-force range $\lambda$.  As expected, the peak sensitivity is achieved when $\lambda$ approximately matches the distance from Jupiter to the perijove of Juno's orbit, since the fifth force exhibits the strongest effect at that scale. 

In addition to our bound, we also present other existing leading constraints in \cref{fig:results}. Planetary constraints are derived using perehilion precession of planets~\cite{Poddar_2021}. We also show constraints from orbital precession of the planet Mars via its tracking data from the Mars Reconnaissance Orbiter (MRO) spacecraft~\cite{konopliv2011mars}. The Lunar Laser Ranging (LLR) constraints are obtained via accurate measurements of the time of flight of a laser pulse making a round trip between an Earth-based observatory and a retroreflector on the Moon~\cite{Williams_2004,Murphy_2013}. Similarly, LAGEOS (Laser Geometric Environmental Observation Survey) constraints are derived from the Earth-LAGEOS satellite and the LAGEOS-Moon system \cite{1985JGR....90.9221S,1987GeoRL..14..730R}.
We highlight that our constraint is the strongest in the range $8.0 \times 10^6$~m $\leq \lambda \leq 6.2 \times 10^7$~m. This corresponds to a force mediator with a mass around $10^{-14}$ eV. 

In \cref{fig:results}, the strongest constraint from Juno on $\alpha$ is obtained for $\lambda\simeq 10^8$ m $\sim \mathcal{O}(\RJupiter)$. This benchmark value corresponds to an anomalous acceleration towards Jupiter's center around $4 \times 10^{-9}$~m/s$^2$ near the perijove, a few times below the empirical uncertainties of residual acceleration $\sim 10^{-8}$ reported in Ref.~\cite{durante2020jupiter}. This could be compatible with the constraints we report since the extra acceleration induced by the fifth force has a definite dependence on the orbit and impacts the orbit cumulatively. A careful analysis should be able to separate the collective effect (like $\Delta \omega$) due to the fifth force from the noise-like residual acceleration. 
The constraint on $\alpha$ scales as $\lambda^2$ for values much larger than the benchmark value $\lambda\gg 10^8$~m. Below the benchmark, for $\lambda\ll 10^7$~m, the constraint instead scales exponentially as $e^{-\RJupiter/\lambda}$. In between these regimes, at values of $\lambda$ around the point of peak sensitivity, the bound strengthens more slowly due to the effects of Jupiter's volume becoming important.

Although not shown in \cref{fig:results}, we want to comment that the constraints from violation of the weak equivalence principle ~\cite{Berg__2018} could be much stronger. However, the equivalence principle constraints are model dependent and only apply to scenarios violating the equivalence principle, while all the bounds in the figure here including ours could apply to scenarios with or without violating the equivalence principle.

\section{Conclusions}\label{section: conclusions}


The orbital data of spacecraft provides a unique and powerful method to probe deviations from gravitational interactions, i.e., from fifth forces. Fifth forces are new interactions mediated by feebly coupled particles, which arise from well-motivated BSM physics. Such a deviation has a characteristic length scale dictated by the mass of the light mediator field; its effect is thus most apparent when gravity is tested around these scales. In this work, we proposed a limit from the orbital data of the Juno mission. While previous studies have primarily focused on heliocentric or geocentric fifth-force constraints, this work extends the analysis to a Jupiter-centric system, leveraging the very recent precise measurements of Juno's orbital velocity. 

Corrections to the inverse-square law of gravity lead to perturbations of orbital elements. The leading behavior of the fifth force is to induce a spherically symmetric anomalous acceleration, always pointing to the center of Jupiter. Therefore, the fifth force's most prominent collective effect is the precession of the orbital element $\omega$. 
Parameterizing the fifth force in terms of its strength $\alpha$ and characteristic range $\lambda$, we computed the drift of the precession $\langle \Delta \omega\rangle
$ due to the fifth force, accounting also for the finite size of Jupiter. Then, to obtain constraints, we compared this to the uncertainty $\sigma_\omega$ in Juno's precession computed using data-inferred properties of Jupiter's gravitational field~\cite{durante2020jupiter}, which was obtained from 10 of the first 17 perijoves of Juno's mission. 

The resulting constraints on the fifth force parameters were summarized in \cref{fig:results}. As a conservative approach, the fifth-force parameter space is excluded when the induced $\langle \Delta \omega\rangle
$ exceeds $2\sigma_\omega$. Scanning the parameter space of $\alpha$ and $\lambda$, we find that the strongest bound is obtained when $\lambda\sim 10^8$~m, the characteristic scale of the Juno kinematics. When $\lambda$ is optimal, all $\alpha \gtrsim 10^{-9}$ values are excluded. The bound on $\alpha$ weakens with $\lambda^2$ in the limit of $\lambda\gg 10^8$~m. 
Between $10^7-10^8$~m, the effect from Jupiter's volume becomes non-negligible. This makes the bound weaken more slowly than naively expected, resulting in the strongest constraint in this range (when assuming no equivalence principle violation). 

The length-scales most sensitively probed by our analysis are noticeably smaller than those probed by many of the lunar and planetary constraints in the literature. This approach takes advantage of the highly eccentric nature of Juno's orbits, as well as the fact that it passes so close to Jupiter's surface, thus probing smaller scales. We can compare to the LAGEOS constraints \cite{1985JGR....90.9221S,1987GeoRL..14..730R}, which arise from man-made geocentric satellites, and thus correspond to similarly small length scales. However, one can see that the precision provided by Juno data improves upon the sensitivity of those constraints considerably.

Our approach displays the power of the Juno data in probing fifth forces, yet the constraints we set can likely be improved further. We do not expect this limit to be the final one from the Jovian-centric field. First, in the short future, incorporating more Juno data will help get stronger constraints. Notice that the current analysis only uses less than 20 perijoves while there are already more than 60 of them. 
In addition, one could move beyond our approach by incorporating the effects of the fifth force directly into the simulation 
of Juno's orbit and environment used to fit to the data. 
Finally, looking into the future, a joint analysis involving data from JUICE~\cite{2012EGUGA..1411912P}, Europa Clipper~\cite{2023SSRv..219...30M}, and other future missions like their successors or Tianwen-4 might benefit the fifth force study even more. While their primary targets are the Jovian moons rather than Jupiter itself, their data might still contribute to a more comprehensive understanding of the Jupiter system and reduce the uncertainties~\cite{2024A&A...687A.132M}.
We note that our approach in this work  can also constrain anomalous mass distributions around Jupiter from BSM effects, such as a dark matter subhalo around Jupiter.
Overall, the sensitivity of data from spacecraft in the vicinity of Jupiter for constraining fifth forces and other BSM phenomenology will undoubtedly keep moving forward.

\section*{Acknowledgement}
We thank Daniele Durante for useful correspondence and discussions. PS, IJA, JF and LL are supported by the NASA grant 80NSSC22K081 and the DOE grant DE-SC-0010010. 

\bibliographystyle{JHEP}
\bibliography{ref.bib}
\appendix

\section{Orbital elements}\label{appendix:orbital_elements}

\begin{figure}[h!]
    \centering
    \includegraphics[width=0.7\textwidth]{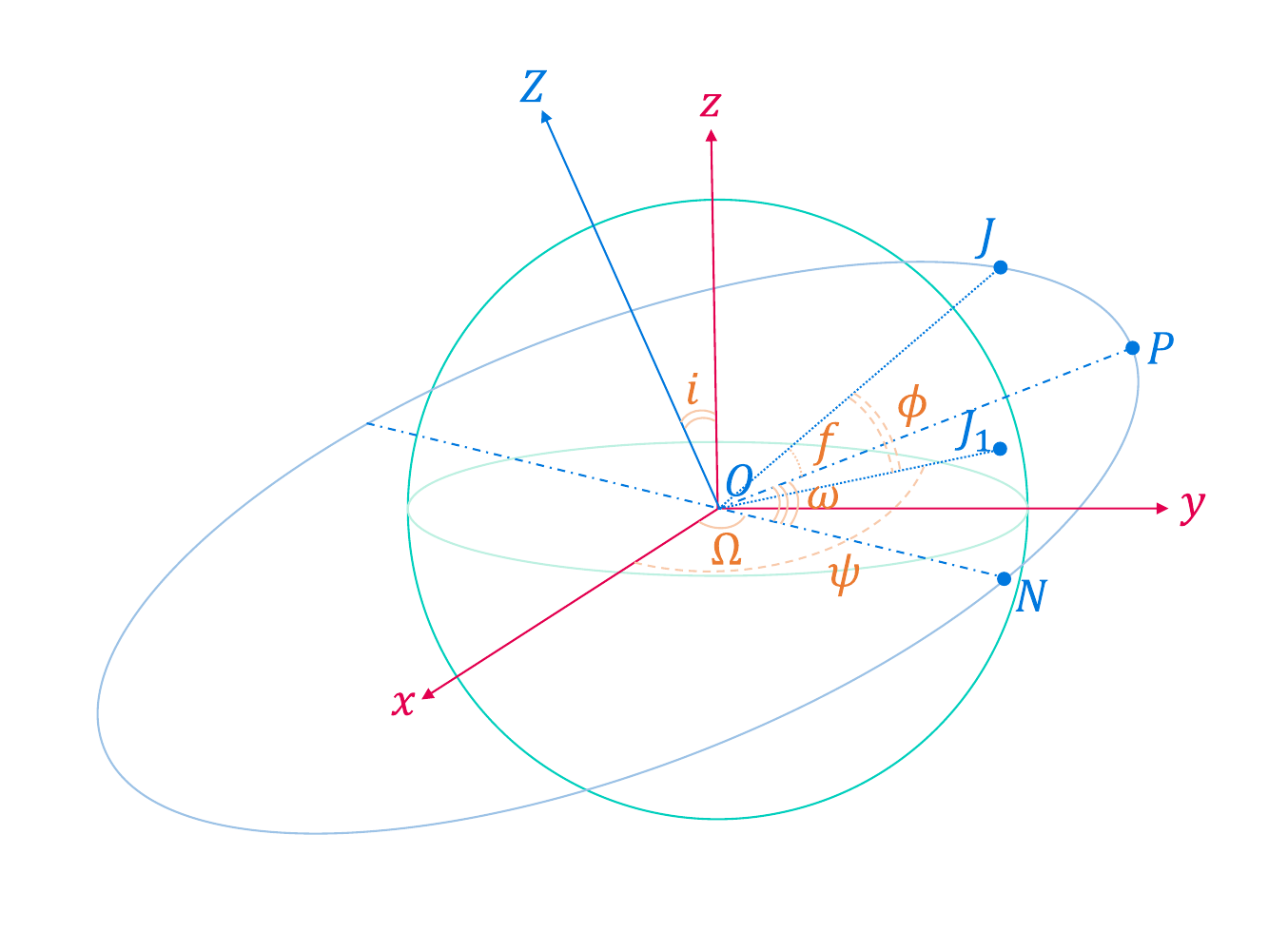}
    \caption{Orbital elements. $O$ is the center of Jupiter, $\pmb{Oz}$ is the axis joining Jupiter's poles, oriented from the south to the north, and $xOy$ is the equatorial plane of Jupiter. $\pmb{Ox}$ is the reference direction and $\pmb{Oy}$ is obtained using right-hand rule. $J$ is Juno's position on its orbital plane and $J_1$ is the projection of $J$ on the equatorial plane. $P$ is the perijove (i.e., periapsis in orbit around Jupiter), $N$ is the ascending node and $\pmb{OZ}$ is the normal vector of the orbital plane. The longitude of ascending node $\Omega$, the inclination angle $i$, and the argument of perijove $\omega$ are the Euler angles. $\psi$ and $\phi$ are the longitude and the latitude of Juno, respectively, and $f$ is its true anomaly.}
    \label{fig:orbital elements}
\end{figure}

As shown in \Fig{fig:orbital elements}, and using the standard astronomical notation, the orbital elements other than the semimajor axis $a$ and eccentricity $e$ are
\begin{itemize}
    \item Longitude of the ascending node
\begin{equation}
    \Omega = ( \pmb{Ox}, \pmb{ON} )~.
\end{equation}
    \item Inclination angle
\begin{equation}
    i = ( \pmb{Oz}, \pmb{OZ} )~.
\end{equation}
    \item Argument of perijove
\begin{equation}
    \omega = ( \pmb{ON}, \pmb{OP} )~.
\end{equation}
    \item True anomaly of Juno
\begin{equation}
    f = ( \pmb{OP}, \pmb{OJ} )~.
\end{equation}
\end{itemize}
Additional parameters include:
\begin{itemize}
    \item Longitude of Juno
\begin{equation}
    \psi = ( \pmb{Ox}, \pmb{OJ_1} )~.
\end{equation}
    \item Latitude of Juno
\begin{equation}
    \phi = ( \pmb{OJ_1}, \pmb{OJ} )~.
\end{equation}
    \item Distance from Juno to Jupiter's center
\begin{equation}
    r = || \pmb{OJ} ||~.
\end{equation}
\end{itemize}

\section{Analytical form of orbital element drifts}\label{app:eqs}

Below, we present expressions for the cycle averaged drift in the precession angle $\langle \Delta \omega \rangle_{J_l}$ due to the non-spherical nature of Jupiter, captured by the zonal harmonic coefficients $J_l$ (see \cref{section: method}). We give also expressions for the $\langle \Delta \Omega \rangle_{J_l}$ and $\langle \Delta i \rangle_{J_l}$ for completeness, though our analysis relies primarily on $\langle \Delta \omega \rangle_{J_l}$.

\subsection{Precession angle $\omega$}
\begin{equation}
      \langle \Delta \omega \rangle_{J_2}=  \frac{3 \pi(3+5 \cos[2 i]) J_2 \RJupiter^2}{4 a^2\left(-1+e^2\right)^2}
\end{equation}

\begin{equation}
      \langle \Delta \omega \rangle_{J_3}=  \frac{3 \pi\left(-1-3 e^2-4 \cos[2 i]+5\left(1+7 e^2\right) \cos[4 i]\right) \csc[i] \times \sin[\omega] J_3 \RJupiter^3}{32 a^3 e\left(-1+e^2\right)^3}
\end{equation}

\bea
\langle \Delta \omega \rangle_{J_4}&= & \frac{1}{512 a^4\left(-1+e^2\right)^4} 15 \pi\bigg(-27\left(4+5 e^2\right)+2\left(-6+5 e^2\right) \cos[2 \omega]\nonumber \\
&& +4 \cos[2 i]\left(-52-63 e^2+2\left(-2+7 e^2\right) \cos[2 \omega]\right)\\
&&+7 \cos[4 i]\left(-28-27 e^2+2\left(2+9 e^2\right) \cos[2 \omega]\right)\bigg) J_4 \RJupiter^4 \nonumber
\eea

\bea
\langle \Delta \omega \rangle_{J_5}&=& 
 \frac{1}{1024 a^5 e\left(-1+e^2\right)^5} 15 \pi\bigg(\left(4+41 e^2+18 e^4\right)(2 \sin[i]+7(\sin[3 i]+3 \sin[5 i])) \sin[\omega]  \nonumber\\
&&+28 e^2\left(1+2 e^2\right)(7+9 \cos[2 i]) \sin[i]^3 \sin[3 \omega]\\
&&-e^2 \cos[i]
\Big[\left(4+3 e^2\right)(2 \cos[i]+21(\cos[3 i]+5 \cos[5 i])) \csc[i] \times \sin[\omega]\nonumber\\
&&+7 e^2(2 \sin[2 i]+15 \sin[4 i]) \sin[3 \omega]\Big]\bigg) J_5 \RJupiter^5\nonumber
\eea

\bea
\langle \Delta \omega \rangle_{J_6}&= & \frac{1}{32768 a^6\left(-1+e^2\right)^6} 105 \pi\bigg(2256 \cos[4 i]+2376 \cos[6 i] \\
&& +5\left(\left(472+1940 e^2+675 e^4\right) \cos[2 i]+3 e^2\left(2\left(292+99 e^2\right) \cos[4 i]+11\left(44+13 e^2\right) \cos[6 i]\right)\right)\nonumber\\
&& -5\Big[10 e^2\left(6+7 e^2\right)+\left(-68+254 e^2+195 e^4\right) \cos[2 i]+6\left(-4+102 e^2+55 e^4\right) \cos[4 i]\nonumber \\
&&+33\left(4+34 e^2+13 e^4\right) \cos[6 i]\Big] \cos[2 \omega]+50\left(24+100 e^2+35 e^4+4 \cos[2 \omega]\right) \nonumber\\
&& -6 e^2\left(-28+45 e^2+4\left(-4+33 e^2\right) \cos[2 i]+11\left(4+13 e^2\right) \cos[4 i]\right) \cos[4 \omega] \sin[i]^2\bigg) J_6 \RJupiter^6\nonumber
\eea

\bea
\langle \Delta \omega \rangle_{J_7}&= & \frac{1}{524288 a^7 e\left(-1+e^2\right)^7} 21 \pi \sin[i]^3 \\
& &\times\bigg(-5\Big[25\left(8+148 e^2+205 e^4+35 e^6\right)+\left(448+6592 e^2+7240 e^4+900 e^6\right) \cos[2 i]\nonumber\\
&&+12\left(56-5 e^2\left(-76+41 e^2+33 e^4\right)\right)\cos[4 i]-132\left(-16-80 e^2+65 e^4\left(2+e^2\right)\right) \cos[6 i]\nonumber\\
&&-429\left(8+212 e^2+365 e^4+75 e^6\right) \cos[8 i]\Big]\csc[i]^4 \sin[\omega]\nonumber\\
&& +30 e^2\Big[14\left(15+7 e^2\right)\left(-8+27 e^2\right)+\left(-2280+13687 e^2+6237 e^4\right) \cos[2 i] \nonumber \\
&& +22\left(24+967 e^2+351 e^4\right) \cos[4 i]+143\left(24+151 e^2+45 e^4\right) \cos[6 i]\Big] \csc[i]^2 \sin[3 \omega]\nonumber \\
&& +264 e^4\left(-45+77 e^2+4\left(-5+52 e^2\right) \cos[2 i]+65\left(1+3 e^2\right) \cos[4 i]\right) \sin[5 \omega]\bigg) J_7 \RJupiter^7\nonumber
\eea

\bea
\langle \Delta \omega \rangle_{J_8}&=& \frac{1}{16777216 a^8\left(-1+e^2\right)^8}  63 \pi\\
&&\times \bigg(5 \Big[-1225\left(192+35 e^2\left(48+56 e^2+9 e^4\right)\right)\nonumber\\
&&-280\left(1664+7 e^2\left(2064+2400 e^2+385 e^4\right)\right) \cos[2 i]\nonumber \\
&& -308\left(1472+7 e^2\left(1776+2040 e^2+325 e^4\right)\right) \cos[4 i]\nonumber\\
&&-3432\left(128+7 e^2\left(144+160 e^2+25 e^4\right)\right)\cos[6 i] \nonumber\\
&& -715\left(704+7 e^2\left(624+600 e^2+85 e^4\right)\right) \cos[8 i]\Big] \nonumber\\
&& +70\Big[35\left(-96+208 e^2+950 e^4+225 e^6\right)+16\left(-384+1648 e^2+5160 e^4+1155 e^6\right) \cos[2 i] \nonumber\\
&& +44\left(-96+1360 e^2+2870 e^4+585 e^6\right) \cos[4 i]+2288 e^2\left(48+80 e^2+15 e^4\right) \cos[6 i] \nonumber\\
&& +143\left(96+1328 e^2+1610 e^4+255 e^6\right) \cos[8 i]\Big] \cos[2 \omega] \nonumber\\
&& +616 e^2\Big[6\left(-280+944 e^2+363 e^4\right)+\left(-1960+14128 e^2+4797 e^4\right) \cos[2 i] \nonumber\\
&& +26\left(40+688 e^2+195 e^4\right) \cos[4 i]+65\left(40+208 e^2+51 e^4\right) \cos[6 i]\Big] \cos[4 \omega] \sin[i]^2 \nonumber\\
&&+4576 e^4\Big[-22+39 e^2+4\left(-2+25 e^2\right) \cos[2 i]+5\left(6+17 e^2\right) \cos[4 i]\Big] \cos[6 \omega] \sin[i]{ }^4\bigg) J_8 \RJupiter^8\nonumber
\eea

\bea
\langle \Delta \omega \rangle_{J_9}&=& \frac{1}{67108864 a^9 e \left(-1+e^2\right)^9}  45 \pi \sin[i]^5\\
&& \times\bigg(7 \Big[98\left(64+1968 e^2+5768 e^4+3325 e^6+315 e^8\right)\nonumber\\
&&+14\left(960+25808 e^2+67032 e^4+33635 e^6+2695 e^8\right)\cos[2 i]\nonumber\\
&& -88\left(-192-3344 e^2-3864 e^4+1225 e^6+455 e^8\right) \cos[4 i] \nonumber\\
&& -429\left(-64+7 e^2\left(-80+232 e^2+405 e^4+65 e^6\right)\right) \cos[6 i] \nonumber\\
&& -286\left(-320+7 e^2\left(-528+488 e^2+1465 e^4+255 e^6\right)\right) \cos[8 i] \nonumber\\
&& -2431\left(64+2608 e^2+9128 e^4+6125 e^6+665 e^8\right) \cos[10 i]\Big] \csc[i]^6 \sin[\omega] \nonumber\\
&& -4312 e^2\Big[21\left(-80+372 e^2+705 e^4+121 e^6\right)+28\left(-96+712 e^2+1178 e^4+195 e^6\right) \cos[2 i] \nonumber\\
&& +52\left(-16+580 e^2+773 e^4+119 e^6\right) \cos[4 i]+52\left(32+744 e^2+834 e^4+119 e^6\right) \cos[6 i] \nonumber\\
&& +221\left(16+156 e^2+147 e^4+19 e^6\right) \cos[8 i]\Big] \csc[i]^4 \sin[3 \omega]\nonumber\\
&& -32032 e^4\Big[22\left(-12+43 e^2+13 e^4\right)+\left(-276+2213 e^2+605 e^4\right) \cos[2 i]\nonumber\\
&& +\left(200+2462 e^2+578 e^4\right) \cos[4 i]+17\left(20+91 e^2+19 e^4\right) \cos[6 i]\Big] \csc[i]^2 \sin[5 \omega] \nonumber\\
&& -4576 e^6\Big[-91+165 e^2+4\left(-7+102 e^2\right) \cos[2 i]+17\left(7+19 e^2\right) \cos[4 i]\Big] \sin[7 \omega]\bigg)J_9 \RJupiter^9\nonumber
\eea

\bea
\langle \Delta \omega \rangle_{J_{10}}&=& \frac{1}{2147483648 a^{10}\left(-1+e^2\right)^{10}} 495 \pi \\
&&\times \bigg(7 \Big[23814\left(128+3 e^2\left(640+1568 e^2+840 e^4+77 e^6\right)\right)\nonumber\\
&&+294\left(20608+307584 e^2+751968 e^4+402360 e^6+36855 e^8\right) \cos[2 i]\nonumber\\
&&+312\left(19072+279936 e^2+679392 e^4+362040 e^6+33075 e^8\right) \cos[4 i] \nonumber\\
&& +1053\left(5504+21 e^2\left(3712+8864 e^2+4680 e^4+425 e^6\right)\right) \cos[6 i] \nonumber\\
&& +442\left(12928+21 e^2\left(8064+18528 e^2+9560 e^4+855 e^6\right)\right) \cos[8 i]\nonumber\\
&& +4199\left(1664+3 e^2\left(5760+7 e^2\left(1632+760 e^2+63 e^4\right)\right)\right) \cos[10 i]\Big)\nonumber\\
&& -84\Big[294\left(-192+544 e^2+6020 e^4+4802 e^6+539 e^8\right)\nonumber\\
&&+98\left(-1088+5344 e^2+42028 e^4+32158 e^6+3549 e^8\right)\cos[2 i]\nonumber\\
&& +104\left(-832+10720 e^2+55356 e^4+39102 e^6+4165 e^8\right) \cos[4 i] \nonumber\\
&& +13\left(-3648+156640 e^2+626444 e^4+411278 e^6+42245 e^8\right) \cos[6 i] \nonumber\\
&& +442\left(64+7328 e^2+24276 e^4+14826 e^6+1463 e^8\right) \cos[8 i] \nonumber\\
&& +4199\left(64+1312 e^2+3220 e^4+1666 e^6+147 e^8\right) \cos[10 i]\Big] \cos[2 \omega] \nonumber\\
&& -8736 e^2\Big[33\left(-224+1208 e^2+1600 e^4+221 e^6\right)\nonumber\\
&&+24\left(-448+3896 e^2+4680 e^4+627 e^6\right) \cos[2 i] \nonumber\\
&& +4\left(-224+30040 e^2+30432 e^4+3825 e^6\right) \cos[4 i]\nonumber\\
&&+136\left(64+920 e^2+808 e^4+95 e^6\right) \cos[6 i] \nonumber\\
&& +323\left(32+248 e^2+192 e^4+21 e^6\right) \cos[8 i]\Big] \cos[4 \omega] \sin[i]^2\nonumber\\
&& -1248 e^4\Big[26\left(-924+3418 e^2+855 e^4\right)+\left(-23100+199186 e^2+45747 e^4\right) \cos[2 i]\nonumber \\
&& +238\left(84+850 e^2+171 e^4\right) \cos[4 i]+323\left(84+346 e^2+63 e^4\right) \cos[6 i]\Big] \cos[6 \omega] \sin[i]^4\nonumber \\
&& -7072 e^6\Big[-120+221 e^2+4\left(-8+133 e^2\right) \cos[2 i]\nonumber\\
&&+19\left(8+21 e^2\right) \cos[4 i]\Big] \cos[8 \omega] \sin[i]^6\bigg) J_{10} \RJupiter^{10} \nonumber
\eea

\subsection{Inclination angle $i$}

\begin{equation}
\langle \Delta i \rangle_{J_2}= 0
\end{equation}

\begin{equation}
\langle \Delta i \rangle_{J_3}= -\frac{3 e \pi \cos[i](3+5 \cos[2 i]) \cos[\omega] J_3 \RJupiter^3}{8 a^3\left(-1+e^2\right)^3}
\end{equation}

\begin{equation}
    \langle \Delta i \rangle_{J_4}=\frac{15 e^2 \pi(5+7 \cos[2 i]) \sin[2 i] \times \sin[2 \omega] J_4 \RJupiter^4}{64 a^4\left(-1+e^2\right)^4}
\end{equation}

\bea
    \langle \Delta i \rangle_{J_5}&=&\frac{1}{512 a^5\left(-1+e^2\right)^5}\\
    &&\times 15 e \pi \cos[i]\Big[\left(4+3 e^2\right)(15+28 \cos[2 i]+21 \cos[4 i]) \cos[\omega]\nonumber\\
    &&+14 e^2(7+9 \cos[2 i]) \cos[3 \omega] \sin[i]^2\Big] J_5 \RJupiter^5\nonumber
    \eea

\bea
 \langle \Delta i \rangle_{J_6}&=& -\frac{1}{8192 a^6\left(-1+e^2\right)^6} 105 e^2 \pi \\
&&\times \Big[5\left(2+e^2\right)(35+60 \cos[2 i]+33 \cos[4 i])\nonumber\\
&&+24 e^2(9+11 \cos[2 i]) \cos[2 \omega] \sin[i]^2\Big] \sin[2 i] \sin[2 \omega] J_6 \RJupiter^6\nonumber
\eea

\bea
 \langle \Delta i \rangle_{J_7}&=& \frac{1}{131072 a^7\left(-1+e^2\right)^7} \\
 &&\times 105 e \pi \cos[i]\Big[-\left(8+5 e^2\left(4+e^2\right)\right)(350+675 \cos[2 i]+594 \cos[4 i]+429 \cos[6 i]) \cos[\omega] \nonumber\\
&& -18 e^2\left(8+3 e^2\right)(189+308 \cos[2 i]+143 \cos[4 i]) \cos[3 \omega] \sin[i]^2\nonumber\\
&&-264 e^4(11+13 \cos[2 i]) \cos[5 \omega] \sin[i]^4\Big] J_7 \RJupiter^7 \nonumber
\eea

\bea
 \langle \Delta i \rangle_{J_8}&=& \frac{1}{1048576 a^8\left(-1+e^2\right)^8}63 e^2 \pi \cos[i] \times \sin[i] \\
&& \times\bigg(35\left(48+80 e^2+15 e^4\right)(210+385 \cos[2 i]+286 \cos[4 i]+143 \cos[6 i]) \sin[2 \omega] \nonumber\\
&& +88 e^2 \sin[i]^2\Big[7\left(10+3 e^2\right)(99+156 \cos[2 i]+65 \cos[4 i]) \sin[4 \omega]\nonumber\\
&&+26 e^2(13+15 \cos[2 i]) \sin[i]^2 \sin[6 \omega]\Big]\bigg) J_8 \RJupiter^8\nonumber
\eea

\bea
 \langle \Delta i \rangle_{J_9}&=& \frac{1}{33554432 a^9\left(-1+e^2\right)^9} \\
&& \times 315 e \pi \bigg(\Big[64+7 e^2\left(48+5 e^2\left(8+e^2\right)\right)\Big]\Big[8722 \cos[i]\nonumber\\
&&+11(756 \cos[3 i]+676 \cos[5 i]+533 \cos[7 i]+221 \cos[9 i])\Big] \cos[\omega]\nonumber \\
&& +1232 e^2\left(16+20 e^2+3 e^4\right) \cos[i](462+819 \cos[2 i]+546 \cos[4 i]+221 \cos[6 i]) \cos[3 \omega] \sin[i]^2\nonumber \\
&& +4576 e^4\left(4+e^2\right)(506 \cos[i]+305 \cos[3 i]+85 \cos[5 i]) \cos[5 \omega] \sin[i]^4\nonumber \\
&& +4576 e^6(47 \cos[i]+17 \cos[3 i]) \cos[7 \omega] \sin[i]^6\bigg) J_9 \RJupiter^9 \nonumber
\eea

\bea 
\langle \Delta i \rangle_{J_{10}}&=& \frac{1}{67108864 a^{10}\left(-1+e^2\right)^{10}} 495 e^2 \pi \cos[i] \sin[i] \\
&&\times \bigg(-21\left(32+7 e^2\left(2+e^2\right)\left(8+e^2\right)\right)\Big[8085+15288 \cos[2 i]+12740 \cos[4 i]\nonumber\\
&&+8840 \cos[6 i]+4199 \cos[8 i]\Big] \sin[2 \omega] \nonumber\\
&& +104 e^2 \sin[i]^2\Big[-42\left(8+8 e^2+e^4\right)(858+1485 \cos[2 i]+918 \cos[4 i]+323 \cos[6 i]) \sin[4 \omega]\nonumber\\
&&-9 e^2\left(14+3 e^2\right)(585+884 \cos[2 i]+323 \cos[4 i]) \sin[i]^2 \sin[6 \omega]\nonumber\\
&&-68 e^4(17+19 \cos[2 i]) \sin[i]^4 \sin[8 \omega]\Big]\bigg) J_{10} \RJupiter^{10}\nonumber
\eea

\subsection{Longitude of the ascending node $\Omega$}

\begin{equation}
\langle \Delta \Omega \rangle_{J_2}= -\frac{3 \pi \cos[i] J_2 \RJupiter^2}{ a^2\left(-1+e^2\right)^2}
\end{equation}

\begin{equation}
\langle \Delta \Omega \rangle_{J_3}=\frac{3 e \pi(7-15 \cos[2 i]) \cot[i] \times \sin[\omega] J_3 \RJupiter^3}{8 a^3\left(-1+e^2\right)^3}
\end{equation}

\begin{equation}
\langle \Delta \Omega \rangle_{J_4}=\frac{15 \pi \cos[i]\left(\left(2+3 e^2\right)(1+7 \cos[2 i])+2 e^2(1-7 \cos[2 i]) \cos[2 \omega]\right) J_4 \RJupiter^4}{32 a^4\left(-1+e^2\right)^4}
\end{equation}

\bea
\langle \Delta \Omega \rangle_{J_5}&=&\frac{1}{1024 a^5\left(-1+e^2\right)^5}\\
&&\times 15 e \pi \Big[(\left(4+3 e^2\right)(2 \cos[i]+21(\cos[3 i]+5 \cos[5 i])) \csc[i] \times \sin[\omega]\nonumber\\
&&+7 e^2(2 \sin[2 i]+15 \sin[4 i]) \sin[3 \omega]\Big] J_5 \RJupiter^5\nonumber
\eea

\bea
\langle \Delta \Omega \rangle_{J_6}&=& \frac{1}{4096 a^6\left(-1+e^2\right)^6}\\
&&\times 105 \pi \cos[i]\Big[-\left(\left(8+40 e^2+15 e^4\right)(19+12 \cos[2 i]+33 \cos[4 i])\right)\nonumber \\
&&+5 e^2\left(2+e^2\right)(41-12 \cos[2 i]+99 \cos[4 i]) \cos[2 \omega]\nonumber\\
&&+6 e^4(7+33 \cos[2 i]) \cos[4 \omega] \sin[i]^2\Big] J_6 \RJupiter^6\nonumber
\eea

\bea
\langle \Delta \Omega \rangle_{J_7}&=& \frac{1}{262144 a^7\left(-1+e^2\right)^7} 21 e \pi \Big[-5\left(8+5 e^2\left(4+e^2\right)\right)\\
&& (25 \cos[i]+243 \cos[3 i]+825 \cos[5 i]+3003 \cos[7 i])  \csc[i] \sin[\omega]\nonumber\\
&&-15 e^2\left(8+3 e^2\right)(45 \sin[2 i]+396 \sin[4 i]+1001 \sin[6 i])\sin[3 \omega] \nonumber\\
&& -528 e^4 \cos[i](29+91 \cos[2 i]) \sin[i]^3 \sin[5 \omega]\Big] J_7 \RJupiter^7\nonumber
\eea

\bea
\langle \Delta \Omega \rangle_{J_8}&=& \frac{1}{524288 a^8\left(-1+e^2\right)^8} 63 \pi \cos[i] \\
&& \times\Big[5\left(16+7 e^2\left(24+5 e^2\left(6+e^2\right)\right)\right)(178+869 \cos[2 i]+286 \cos[4 i]+715 \cos[6 i])\nonumber\\
&& -70 e^2\left(48+80 e^2+15 e^4\right)(-8+121 \cos[2 i]+143 \cos[6 i]) \cos[2 \omega]\nonumber \\
&& -616 e^4\left(10+3 e^2\right)(43+52 \cos[2 i]+65 \cos[4 i]) \cos[4 \omega] \sin[i]^2 \nonumber\\
&& -4576 e^6(2+5 \cos[2 i]) \cos[6 \omega] \sin[i]^4\Big] J_8 \RJupiter^8\nonumber
\eea

\bea
\langle \Delta \Omega \rangle_{J_9}&=& \frac{1}{33554432 a^9\left(-1+e^2\right)^9} 45 e \pi\Big[7\left(64+7 e^2\left(48+5 e^2\left(8+e^2\right)\right)\right) \\
&& \times(98 \cos[i]+11(84 \cos[3 i]+260 \cos[5 i]+637 \cos[7 i]+1989 \cos[9 i])) \csc[i] \times \sin[\omega] \nonumber\\
&& +2156 e^2\left(16+20 e^2+3 e^4\right)(14 \sin[2i]+130(\sin[4 i]+3 \sin[6 i])+663 \sin[8 i]) \sin[3 \omega] \nonumber\\
&& +32032 e^4\left(4+e^2\right)(418 \cos[i]+325 \cos[3 i]+153 \cos[5 i]) \sin[i]^3 \sin[5 \omega] \nonumber\\
&& +9152 e^6 \cos[i](71+153 \cos[2 i]) \sin[i]^5 \sin[7 \omega]\Big] J_9 \RJupiter^9\nonumber
\eea

\bea
\langle \Delta \Omega \rangle_{J_{10}}&=& \frac{1}{268435456 a^{10}\left(-1+e^2\right)^{10}}495 \pi \cos[i] \\
&& \bigg(-7\left(128+2304 e^2+6048 e^4+3360 e^6+315 e^8\right)(2773+2392 \cos[2 i] \nonumber\\
&& +5252 \cos[4 i]+1768 \cos[6 i]+4199 \cos[8 i]) \nonumber\\
&& +84 e^2\left(32+7 e^2\left(2+e^2\right)\left(8+e^2\right)\right)\Big[8553-936 \cos[2 i]+18772 \cos[4 i]\nonumber\\
&&+1768 \cos[6 i]+20995 \cos[8 i]\Big] \cos[2 \omega] \nonumber\\
&& +8736 e^4\left(8+8 e^2+e^4\right)(1098+2721 \cos[2 i]+1734 \cos[4 i]+1615 \cos[6 i]) \cos[4 \omega] \sin[i]^2 \nonumber\\
&& +1248 e^6\left(14+3 e^2\right)(1517+2244 \cos[2 i]+1615 \cos[4 i]) \cos[6 \omega] \sin[i]^4 \nonumber\\
&&+7072 e^8(49+95 \cos[2 i]) \cos[8 \omega] \sin[i]^6\bigg) J_{10} \RJupiter^{10}\nonumber 
\eea

\end{document}